\documentstyle[12pt]{article}
\makeatletter
\newdimen\normalarrayskip              
\newdimen\minarrayskip                 
\normalarrayskip\baselineskip
\minarrayskip\jot
\newif\ifold             \oldtrue            
\def\arraymode{\ifold\relax\else\displaystyle\fi} 
\def\eqnumphantom{\phantom{(\theequation)}}     
\def\@arrayskip{\ifold\baselineskip\z@\lineskip\z@
     \else
     \baselineskip\minarrayskip\lineskip2\minarrayskip\fi}
\def\@arrayclassz{\ifcase \@lastchclass \@acolampacol \or
\@ampacol \or \or \or \@addamp \or
   \@acolampacol \or \@firstampfalse \@acol \fi
\edef\@preamble{\@preamble
  \ifcase \@chnum
     \hfil$\relax\arraymode\@sharp$\hfil
     \or $\relax\arraymode\@sharp$\hfil
     \or \hfil$\relax\arraymode\@sharp$\fi}}
\def\@array[#1]#2{\setbox\@arstrutbox=\hbox{\vrule
     height\arraystretch \ht\strutbox
     depth\arraystretch \dp\strutbox
     width\z@}\@mkpream{#2}\edef\@preamble{\halign \noexpand\@halignto
\bgroup \tabskip\z@ \@arstrut \@preamble \tabskip\z@ \cr}%
\let\@startpbox\@@startpbox \let\@endpbox\@@endpbox
  \if #1t\vtop \else \if#1b\vbox \else \vcenter \fi\fi
  \bgroup \let\par\relax
  \let\@sharp##\let\protect\relax
  \@arrayskip\@preamble}
\def\eqnarray{\stepcounter{equation}%
              \let\@currentlabel=\theequation
              \global\@eqnswtrue
              \global\@eqcnt\z@
              \tabskip\@centering
              \let\\=\@eqncr
              $$%
 \halign to \displaywidth\bgroup
    \eqnumphantom\@eqnsel\hskip\@centering
    $\displaystyle \tabskip\z@ {##}$%
    &\global\@eqcnt\@ne \hskip 2\arraycolsep
         $\displaystyle\arraymode{##}$\hfil
    &\global\@eqcnt\tw@ \hskip 2\arraycolsep
         $\displaystyle\tabskip\z@{##}$\hfil
         \tabskip\@centering
    &{##}\tabskip\z@\cr}
\makeatother
\catcode`\@=11
\ifcase\@ptsize
 \font\tenmsa=msam10
 \font\sevenmsa=msam7
 \font\fivemsa=msam5
 \font\tenmsb=msbm10
 \font\sevenmsb=msbm7
 \font\fivemsb=msbm5
 \font\teneu=eufm10
 \font\seveneu=eufm7
 \font\fiveeu=eufm5
 \font\tenib=cmmib10
 \font\sevenib=cmmib7
 \font\fiveib=cmmib5
\or
 \font\tenmsa=msam10 scaled \magstephalf
 \font\sevenmsa=msam7 scaled \magstephalf
 \font\fivemsa=msam5 scaled \magstephalf
 \font\tenmsb=msbm10 scaled \magstephalf
 \font\sevenmsb=msbm7 scaled \magstephalf
 \font\fivemsb=msbm5  scaled \magstephalf
 \font\teneu=eufm10  scaled \magstephalf
 \font\seveneu=eufm7  scaled \magstephalf
 \font\fiveeu=eufm5   scaled \magstephalf
 \font\tenib=cmmib10  scaled \magstephalf
 \font\sevenib=cmmib7  scaled \magstephalf
 \font\fiveib=cmmib5   scaled \magstephalf
\or
 \font\tenmsa=msam10 scaled \magstep1
 \font\sevenmsa=msam7 scaled \magstep1
 \font\fivemsa=msam5  scaled \magstep1
 \font\tenmsb=msbm10 scaled \magstep1
 \font\sevenmsb=msbm7 scaled \magstep1
 \font\fivemsb=msbm5  scaled \magstep1
 \font\teneu=eufm10   scaled \magstep1
 \font\seveneu=eufm7 scaled \magstep1
 \font\fiveeu=eufm5 scaled \magstep1
 \font\tenib=cmmib10     scaled \magstep1
 \font\sevenib=cmmib7   scaled \magstep1
 \font\fiveib=cmmib5   scaled \magstep1
\fi
\newfam\msafam
\textfont\msafam=\tenmsa  \scriptfont\msafam=\sevenmsa
  \scriptscriptfont\msafam=\fivemsa
\newfam\msbfam
\textfont\msbfam=\tenmsb  \scriptfont\msbfam=\sevenmsb
  \scriptscriptfont\msbfam=\fivemsb
\def\Bbb{\ifmmode\let\next\Bbb@\else
 \def\next{\errmessage{Use \string\Bbb\space only in math mode}}\fi\next}
\def\Bbb@#1{{\Bbb@@{#1}}}
\def\Bbb@@#1{\fam\msbfam#1}
\newfam\eufam
\textfont\eufam=\teneu  \scriptfont\eufam=\seveneu
  \scriptscriptfont\eufam=\fiveeu
\def\frak{\ifmmode\let\next\frak@\else
 \def\next{\errmessage{Use \string\frak\space only in math mode}}\fi\next}
\def\frak@#1{{\frak@@{#1}}}
\def\frak@@#1{\fam\eufam#1}
\newfam\ibfam
\textfont\ibfam=\tenib  \scriptfont\ibfam=\sevenib
  \scriptscriptfont\ibfam=\fiveib
\def\bold{\ifmmode\let\next\bold@\else
\def\next{\errmessage{Use \string\bold\space only in math mode}}\fi\next}
\def\bold@#1{{\bold@@{#1}}}
\def\bold@@#1{\fam\ibfam#1}
\catcode`\@=12
\begin{document}
\def\bea{\begin{eqnarray}}
\def\eea{\end{eqnarray}}
\def\nn{\nonumber}
\def\ZZ{\Bbb{Z}}
\def\CC{\Bbb{C}}
\def\DD{\Bbb{D}}
\def\dis#1{\displaystyle{#1}}
\def\stackreb#1#2{\mathrel{\mathop{#2}\limits_{#1}}}
\def\res#1{\stackreb{#1}{\rm res}}
\def\ep{\varepsilon}

\begin{center}
\hfill DFTUZ/94/07\\
\hfill hep-th/9405018\\
\bigskip\bigskip
{\Large  Relations Between Hyperelliptic Integrals}\\
\bigskip
\bigskip
{\large S. Pakuliak}\footnote{E-mail:
pakuliak@cc.unizar.es}
\footnote{On leave of absence from the ITP, Kiev 252143, Ukraine}
\ \ {\large and}\ \
{\large A. Perelomov}\footnote{E-mail:
perelomo@pib1.physik.uni-bonn.de}
\footnote{On leave of absence from the ITEP, Moscow, Russia}\\
\bigskip
{\it Departamento de F\'\i sica Te\'orica\\
Facultad de Ciencias\\
Universidad de Zaragoza\\
50009 Zaragoza, Spain}\\
\bigskip
\end{center}
\begin{abstract}
A simple property of the integrals over
the hyperelliptic surfaces of arbitrary genus is observed. Namely,  the
derivatives of these integrals with respect to the branching
points are given by the linear combination of the same
integrals. We check that this property is responsible
for the solution  to the level zero Knizhnik-Zamolodchikov
equation given in terms of hyperelliptic integrals.
\end{abstract}

\section{Introduction}
\setcounter{footnote}{0}

The starting point of our investigation is the observation due
to F.A. Smirnov \cite{Smpr92,Sm93} that the integral
representation \cite{CF87,SV91,DJMM90} for the solutions
to the Knizhik-Zamolod\-chi\-kov equation restricted to
the level zero and associated with affine
$\widehat{\frak{sl}}_2$ algebra can be rewritten as the determinant
of the matrix  having  second kind
hyperelliptic integrals as its elements.
The KZ equation in this case is the
system of first order differential equations for the
multicomponent function
$f_{\ep_1,\ldots,\ep_n}(\lambda_1,\ldots,\lambda_n)\in \CC^2\otimes
\ldots\otimes\CC^2$,
$\ep_i=\pm$, $\lambda_i\in\CC$
\begin{equation}
\left({\partial\over \partial\lambda_j}-{1\over4}
\sum_{i\neq j}{\sigma^a_i\otimes\sigma^a_j\over
\lambda_j-\lambda_i} \right)f(\lambda_1,\ldots,\lambda_n)=0,
\label{KZ0}
\end{equation}
where the operators $\sigma_i^a=1\otimes\ldots\otimes1\otimes
\sigma^a\otimes1\otimes\ldots\otimes1$ act as the Pauli matrices in the $i$th
two-dimensional vector space $\CC^2$.
System  (\ref{KZ0}) is obviously splitted to the subsystems for
the functions
$f_{\ep_1,\ldots,\ep_n}(\lambda_1,\ldots,\lambda_n)$ with
$\#\{\ep_i=+\}=l$ and $\#\{\ep_i=-\}=m$  being fixed. These
subsystems correspond to the fixed total spin $(l-m)/2$.

The question that we address in this letter is why the
determinant combination of the full hyperelliptic integrals,
which are complicated transcendental functions of the branching
points $\lambda_i$, satisfy a simple differential equation with
rational coefficients.

We found the explanation to this phenomena, namely to the fact that
the derivatives of the full hyperelliptic integral with respect to the
branching points can be written as a linear combination of the
same integrals. The proof is just the application of the
integration by part rule to some third kind integral
(\ref{int9}).

Since the hyperelliptic integrals appear in many problems
of the mathematical physics we hope that the formulas of this note
will be useful in application to such problems. Let us formulate the
main result of this letter.

Consider the Riemann hyperelliptic surface of genus $g$ given in $\CC^2$
by algebraic relation
\begin{equation}
y^2=\prod_{k=1}^{n}(x-\lambda_k),
\label{sur2}
\end{equation}
where $n=2g+1$ or $n=2g+2$ depending on the case whether the infinity
is  the branching point or not. Using rational transformation we
can always place all the branching points in the finite part of the
complex plane, so without loosing the generality we will restrict ourselves
to the case when $n=2g+2$. Define the integrals on this Riemann surface
\begin{equation}
K_j(\lambda)=\int_{\gamma}{x^jdx\over\sqrt{P_{2g+2}(x)}},\quad
 j\in\ZZ_+,  \label{int3}
 \end{equation}
\begin{equation}
P_{2g+2}(x) = \prod_{i=1}^{2g+2} (x-\lambda_k)=\sum_{i=0}^{2g+2}
(-)^i\sigma_i(\lambda)x^{2g+2-i}, \nn
\end{equation}
where $\sigma_i(\lambda)$ are order $i$ homogeneous  and
symmetric functions  of the branching points $\lambda_j$, and
$\gamma$ is an arbitrary closed contour on the surface
that in the case under consideration
can be reduced to the sum of the integrals
between points $\lambda_j$.  We state a
\medskip

\noindent{\bf Proposition.} {\sl The integrals (\ref{int3})
with $j=0,1,\ldots,2g$, as functions of the branching points satisfy the
``minimal''
\footnote{This word will be explained in Sect. 3.}
system of the differential equations of the first order
\bea
{\partial K_j(\lambda)\over \partial\lambda_m}&={1\over2}
\left(\sum_{i=0}^{j-1}\lambda_{m}^{j-i-1}K_i(\lambda)+
{\lambda^j_m\over\hat P^{(m)}_{2g+1}(\lambda_m)}
\sum_{i=0}^{2g}a_{i}(\lambda)K_i(\lambda)\right)\nn\\
a_{i}(\lambda)&=(-)^i\left(\sum_{k\neq m}\hat\sigma^{(m,k)}_{2g-i}
-\sum_{m=0}^{2g-i}(-)^k\lambda_m^k \hat\sigma^{(m)}_{2g-i-m} \right),
\label{for4}
\eea
where  $m=1,\ldots,2g+2$,
$$ \hat P^{(m)}_{2g+1}(x)=\prod_{l=1\atop l\neq m}^{2g+2}(x-\lambda_l)=
\sum_{i=0}^{2g+1}(-)^i\hat\sigma^{(m)}_{i}x^{2g+1-i},$$
and $\hat\sigma^{(m)}_{i}$ are symmetric function of the order $i$ of
the variables  $\lambda_l$, $l=1,\ldots,m-1,m+1,\ldots,2g+2$.
The symmetric functions $\hat\sigma^{(m,k)}_{i}$ are defined
analogously
$$ \hat P^{(m,k)}_{2n-2}(x)=\prod_{l=1\atop l\neq m,k}^{2g}(x-\lambda_l)=
\sum_{i=0}^{2g}(-)^i\hat\sigma^{(m,k)}_{i}x^{2g-i}.$$
}
\medskip

\noindent Using the obvious identity
\begin{equation}
\sum_{k=1\atop k\neq m}^{M}\hat\sigma_j^{(k,m)}= (M-1-j)
\hat\sigma_j^{(m)},\quad j=0,1,\ldots,M-2
\label{sym41}
\end{equation}
we can express the coefficients $a_{i}$ only in terms of symmetric
functions $\hat\sigma^{(m)}_i$
\begin{equation}
a_{i}(\lambda)=(-)^i\left(i\hat\sigma^{(m)}_{2g-i}
-\sum_{k=1}^{2g-i}(-)^k\lambda_m^k \hat\sigma^{(m)}_{2g-i-k} \right).
\label{for42}
\end{equation}

\section{Proof}

To prove the above statement we observe first that the derivative
\begin{equation}
{\partial K_j(\lambda)\over \partial\lambda_m}  =
{1\over2}\int_\gamma {x^jdx\over\sqrt{P_{2g+2}(x)}(x-\lambda_m)}
\label{der5}
\end{equation}
is independent on the contour $\gamma$.
Ended, using the homology symmetry on the Riemann surface
we can always remove the dependence on the variable $\lambda_m$ in the
integration limits. In what follows we will always assume that the
contour $\gamma$ is chosen in such a way that it does not cross the
branching point $\lambda_m$. As result we can conclude that derivative
of $K_j$ with respect to any branching point is given by  integral
(\ref{der5}). Dividing $x^j/(x-\lambda_m)$, we obtain the first term
in (\ref{for4}), and calculation of (\ref{der5}) is reduced to the
calculation of the integral
\begin{equation}
\int_{\gamma}   {dx\over\sqrt{P_{2g+2}(x)}(x-\lambda_m)}.
\label{int6}
\end{equation}
To calculate integral (\ref{int6}) we use the following trick. Let us
again use the fact that the integral over
any closed contour on the Riemann surface $y^2=P_{2g+2}(x)$
can be reduced to the integral between the branchings points.
In this case the integral
\begin{equation}
\int_{\gamma}dx {d\over dx}   {\sqrt{P_{2g+2}(x)}\over
(x-\lambda_m)}=0
\label{int7}
\end{equation}
is identically zero if we again  adjust properly the
contour $\gamma$ in (\ref{int7}).

Calculating the derivative under the integral in (\ref{int7})
we arrive to the relation
\begin{equation}
\int_{\gamma}{\dis{\sum_{k=1\  k\neq m}^{2g+2}
\hat P^{(m,k)}_{2g}(x)}\over \sqrt{P_{2g+2}(x)}}=
\int_{\gamma}
\dis{\hat P^{(m)}_{2g+1}(x)\over \sqrt{P_{2g+2}(x)} (x-\lambda_m)}.
\label{rel8}
\end{equation}
Dividing $\hat P^{(m)}_{2g+1}(x) /(x-\lambda_m)$, we obtain
\begin{equation}
\int_{\gamma}  {dx\over\sqrt{P_{2g+2}(x)}
(x-\lambda_m)}=
{1\over\hat P^{(m)}_{2g+1}(\lambda_m)}
\sum_{i=0}^{2g}a_{i}(\lambda)K_i(\lambda),
\label{int9}
\end{equation}
where coefficients $a_{i}$  are given by  (\ref{for4}). This finishes the
proof of the proposition.

Let us point out that for the elliptic surface the formulas (\ref{for4})
are well known
and can be found in most books on special functions (see for example
\cite{RG80}, formulas 8.123.1-4).

\section{Discussion}

Let us explain the word ``minimal'' used in the formulation of the
proposition. To obtain formulas like (\ref{for4}),
we can start from the integral
$K_j(\lambda)$ for any $j\geq0$ and calculate the derivative of this
integral with respect to any branching point. Then, as we have seen in the
previous section, the minimal number of integrals $K_i$ contained in the
formula for the derivative
is equal to $2g+1$, $0\leq i\leq 2g$.
It explains why we choose this minimal number of integrals to
write down the formulas (\ref{for4}).
This minimal set of integrals consists of
$g$ integrals of the first kind differentials
($x^i/\sqrt{P_{2g+2}(x)}$, $0\leq i\leq g$) with no
singularities at the complex plain, $g$ integrals of the second
kind differentials ($x^i/\sqrt{P_{2g+2}(x)}$, $g+1\leq i\leq
2g$) with pole singularity at infinity points $\infty^\pm$
of the order $i-g+1$ and one third kind integral of
$x^g/\sqrt{P_{2g+2}(x)}$ with logarithmic singularity at the
infinity.

Because of the uniqueness of the third kind integral we can reduce the system
(\ref{for4}).
Let us introduce the integrals of the second kind differentials
which has zero residues at infinity,
\begin{equation}
E_j=K_j-c_jK_g,\quad  j = 0,\ldots,2g,
\label{int10}
\end{equation}
where
\begin{equation}
c_j=\res{x=\infty}{x^j\over\sqrt{P_{2g+2}(x)}}.
\label{res11}
\end{equation}
There is a simple recurrent relation for the symmetric
functions $c_j(\lambda)$ that follows
from\footnote{Equation (\ref{der111})  can be easily obtained
by taking the
derivative of (\ref{res11}).}
\begin{equation}
2{\partial c_j(\lambda)\over\partial\lambda_m} =
\sum_{i=g}^{j-1}\lambda_m^{j-i-1}c_i(\lambda) ,\quad j=g+1,\ldots,2g.
\label{der111}
\end{equation}
Multiplying the left and right sides of (\ref{der111}) by $\lambda_m$ and
summing over $m$, we obtain
\begin{equation}
c_j(\lambda)={1\over2(j-g)}\sum_{i=g}^{j-i}s_{j-i}(\lambda)c_i(\lambda),
\label{rec112}
\end{equation}
where we have used the property of the order $(j-g)$ homogeneous functions
$c_j(\lambda)$ and introduced new symmetric functions
\begin{equation}
s_i(\lambda)=\sum_{m=1}^{2g+2}\lambda_m^{i}.
\label{sym113}
\end{equation}
Using the recurrent relation (\ref{rec112}) with a boundary condition
$c_g=1$,  one can easily calculate the functions $c_j$. For example,
\bea
c_{g+1}&={1\over2}s_1={1\over2}\sigma_1\ ,\nn\\
c_{g+2}&={1\over4}s_2+{1\over8}s_1^2={3\over8}\sigma_1^2
-{1\over2}\sigma_2\ ,\nn\\
c_{g+3}&={1\over6}s_3+{1\over8}s_1s_2+{1\over48}s_1^3
={5\over16}\sigma_1^3-{3\over4}\sigma_1\sigma_2+{1\over2}\sigma_3\ .\nn
\eea

Let us point out that the set of functions $c_j$ forms a new basis
in the
space of homogeneous and symmetric functions of many variables
different from the bases generated by $s_j$ and $\sigma_j$ with
a good property (\ref{der111}).
On the other hand, the symmetric functions $c_j(\lambda)$ are the
coefficients of the expansion of the function $1/\sqrt{P_{2g+2}(x)}$ in
the vicinity of the infinity point. The recurrent relation
(\ref{rec112}) is much more convenient for calculation these
coefficients than the direct expansion of this function.

Now we are in the position to reduce
$2g+1$-dimensional system (\ref{for4}) for the integrals $K_j$
to $2g$ dimensional system for the integrals $E_j$.
\bea
{\partial E_j(\lambda)\over \partial\lambda_m}&={1\over2}
\left(\sum_{i=0}^{j-1}\lambda_{m}^{j-i-1}E_i(\lambda)-
c_j(\lambda)\sum_{i=0}^{g-1}\lambda_{m}^{g-i-1}E_i(\lambda)\right) \nn\\
&+
{\lambda^j_m-c_j(\lambda)\lambda_m^g\over2\hat P^{(m)}_{2g+1}(\lambda_m)}
\sum_{i=0\atop i\neq g}^{2g}a_{i}(\lambda)E_i(\lambda).
\label{newsys114}
\eea

This reduction follows from the nontrivial identity between the
functions $c_j(\lambda)$ and $a_i(\lambda)$
\begin{equation}
\sum_{i=g}^{2g}a_i(\lambda)c_i(\lambda)=0.
\label{ident12}
\end{equation}

It was demonstrated in \cite{Smpr92} that the linear
combination of the determinants composed from  the integrals
$E_i$, $i=g+1,\ldots,2g$,
satisfies the level zero KZ equation (\ref{KZ0}).  The papers
\cite{Smpr92,Sm93} were devoted to the investigation of the
classical limit of quantum KZ equation \cite{FR92}.
The quantum KZ equation is a system of difference
equations which historically originates in the bootstrap
approach in quantum field theory.  F.A. Smirnov developed the systematic
approach to the solutions of this kind difference systems that
was summarized in the book \cite{Smbook}. The quantum or
deformed KZ equation at level zero coincide essentially with
the form factors equations in the completely integrable models
of quantum field theory.
The
solutions to the form factor equations for the models which are
associated with Yangian and $U_q(\widehat{\frak{sl}}_2)$,
$|q|=1$ symmetries  and
restricted to the total spin zero
case
were considered in that book. Recently the
same approach was successfully applied to the construction of
the integral solution to the level 0 deformed KZ equation
associated with $U_q(\widehat{\frak{sl}}_2)$, $|q|<1$ symmetry
and for arbitrary values of the total spin
\cite{JKMQ93}.

One of the  main ingredients of the Smirnov's approach is the
identities between integrals of some meromorphic functions
which are deformed analogues of the holomorphic differentials on
the hyperelliptic Riemann surface. Using these identities and
quasi-periodicity properties of the kernel of the integral
solution, one can solve the deformed KZ equation.

The integral solutions to the classical and quantum KZ
equation have quite
different properties due to the problem of braiding. In the
deformed case the integrands for the solutions are the
meromorphic functions which have infinite number of
simple poles and  essential singularities at infinity that
makes the braiding trivial. In the classical limit these
infinite sequences of the poles are concentrating into the cuts
that leads to complicated braiding which is  served by a
finite-dimensional  quantum groups.  This makes the relation
between the solutions to the deformed and the ordinary KZ equations
quite complicated and only asymptotical (see Ref. \cite{Smpr92} for
the precise treatment). But nevertheless, as we have seen above,
it is possible to find quite simple identities between
hyperelliptic integrals and then to show that the complicated
transcendental solutions to the level zero KZ equation    are
the consequences of these relations.
The integral formulas for the solution to this equation
in the subsector of the total spin zero
were written in \cite{Smpr92}, while those for the
non-zero subsectors can be obtained after the classical limit
from the formulas presented in \cite{JKMQ93}.

\section{Conclusion}

In this note we have addressed to the question why the simple dynamical
systems with rational coefficients like level zero KZ equation
     possess the complicated transcendental
solution \cite{SV91} defined on the hyperelliptic surface. We have found
the simple relation between the hyperelliptic integrals which is
responsible for this phenomena.

To conclude we would like to point out the questions which are very
interesting from our point of view and deserve further investigation.
\medskip

\begin{itemize}
\item Whether is it possible to obtain by the pure analytical tools the
integral formula for the solution of KZ equation at the arbitrary level.
We suppose that the key step to solve this problem is to use the
equivalence between KZ and Calogero systems \cite{Ma92} and a starting point
there should be the ground wave function for the Calogero system
containing the different sort of particles.\\
\item It is an interesting problem to investigate the operator content of the
formulas (\ref{newsys114}) in the sense of the field theory on algebraic
curves. An exhaustive investigation
in this direction was done recently in preprint \cite{FSU93} for
for the field theories on the $\ZZ_n$-symmetric algebraic curves and
in \cite{F93} for the $\DD_n$-symmetric ones.
\item As it is known \cite{Fa73}, the hyperelliptic integrals can
be expressed in terms of the theta constants. It is an interesting question
what relations between the theta constants correspond to
formulas (\ref{for4}).
\end{itemize}
\medskip

\section{Acknowledgments}

The authors would like to acknowledge Departamento de F\'\i sica
Te\'orica of Universidad de Zaragoza where this work was done.
The research of S.P. was supported by Direcci\'on General de
Investigaci\'on Cient\'\i fica y T\'ecnica (Madrid).



\end{document}